\documentclass[12pt, a4paper]{article}
\usepackage[T1]{fontenc}
\usepackage{graphicx}
\usepackage[linesnumbered,ruled,vlined,resetcount,noend]{algorithm2e}
\usepackage{url}
\usepackage{amssymb}

\usepackage{hyperref}
\usepackage{comment}
\usepackage{multicol}
  \usepackage{subcaption}
\usepackage{mathtools}

\usepackage{multirow}
\usepackage{xcolor}
\usepackage{tcolorbox}
\usepackage{amsthm}

\newtheorem{definition}{Definition}[section]
\usepackage{adjustbox}
\usepackage{authblk}
\usepackage{caption}
\usepackage{caption}
\usepackage{subcaption}
\usepackage{amsmath}
\usepackage{amssymb}
\usepackage{float}
\usepackage{graphicx}
\usepackage{pifont}
\usepackage{wrapfig}
\usepackage{enumitem}
\usepackage[noadjust]{cite}
\usepackage{tikz}

\newcounter{protocol}

\usepackage [
n ,
advantage ,
operators,
sets,
adversary ,
landau ,
probability ,
notions, 
logic,
ff,
mm,
primitives,
events,
complexity,
asymptotics ,
keys
]{cryptocode}

\begin{document}
 \newcommand{\moh}[1]{{[\textcolor{purple}{Mohammed: #1}]}}
\pagestyle{plain}

\newcommand{\ccr}{\mathrm{CCR}}
\newcommand{\curk}{\mathrm{k}}
\newcommand{\pcr}{\mathrm{PCR}}
\newcommand{\spcr}{\mathrm{SPCR}}
\newcommand{\pce}{\mathrm{PCE}}
\newcommand{\thresval}{\mathrm{ThresholdValue}}
\newcommand{\chancap}{\mathrm{C}}
\newcommand{\curi}{\mathrm{i}}
\newcommand{\curj}{\mathrm{j}}
\newcommand{\sybil}{\mathsf{Attacker}}
\newcommand{\attacker}{\mathsf{AdV}}
\newcommand{\targetnode}{\mathrm{Target_{node}}}
\newcommand{\sset}{\mathcal{A}}
\newcommand{\pset}{\mathbb{P}}
\newcommand{\maxflow}{\mathsf{Probing}}
\newcommand{\curmax}{\mathrm{currmax}}
\newcommand{\lightning}{\mathsf{LN}}
\newcommand{\Capcitylist}{\mathbb{C}}
\newcommand{\Paymentlist}{\mathcal{P}}
\newcommand{\targetlist}{\mathbb{Target_{list}}}
\newcommand{\Tset}{\mathbb{T}}
\newcommand{\ulock}{\mathsf{SCOOP}}
\newcommand{\minpay}{\mathsf{MinPay}}
\newcommand{\spcrmax}{\mathsf{SPCR~Max}}

\newcommand{\sourceroute}{\mathsf{FindAllPaths}}
\newcommand{\congest}{\mathsf{ChannelCongest}}
\newcommand{\RPO}{\mathit{Payment~
Minimization}}
\newcommand{\PCO}{\mathit{Congestion~Maximization}}
\newcommand{\alice}{\mathsf{Alice}}
\newcommand{\peter}{\mathsf{Peter}}
\newcommand{\carol}{\mathsf{Carol}}
\newcommand{\alexa}{\mathsf{Alexa}}
\newcommand{\megan}{\mathsf{Megan}}
\newcommand{\attackerone}{\mathsf{Attacker1}}
\newcommand{\attackertwo}{\mathsf{Attacker2}}
\newcommand{\attackerthree}{\mathsf{Attacker3}}

\newcommand{\add}{\mathsf{UNION}}
\newcommand{\createhtlc}{\mathsf{CreateHTLC}}
\newcommand{\deletehtlc}{\mathsf{failHTLC}}
\newcommand{\tset}{\mathbb{T}}
\newcommand{\bal}{\mathsf{balance}}

\newcommand{\pcopen}{\mathsf{PC.Open}}

\newcommand{\createsybil}{\mathsf{CreateSybil}}

\newcommand{\wellconnectednodes}{\mathsf{WellConnectedNodes}}
\newcommand{\hopcount}{\mathsf{HopCount}}
\newcommand{\randomnodes}{\mathsf{RandomNodeSelection}}
\newcommand{\selectnode}{\mathsf{SelectNode}}
\newcommand{\runlightningnode}{\mathsf{LNRPC}}
\newcommand{\RWCN}{\mathsf{RandomWellConnectedNodes}}
\newcommand{\STN}{\mathsf{SelectTargetNode}}
\newcommand{\mk}{\mathsf{CreateChannel}}

\newcommand{\findpath}{\mathsf{Congestion \ Attacks}}
\newcommand{\pathcongest}{\mathsf{Congestion \ Attack \ 2}}
\newcommand{\setup}{\mathsf{Setup \ And \ Sybil \ Node \ Creation}}
\newcommand{\updat}{\mathsf{Sybil \ Node \ Generating \ and \ Channel \ Creation}}

\title{$\ulock$: Co\underline{S}t-effective \underline{C}\underline{O}ngesti\underline{O}n Attacks in \underline{P}ayment Channel Networks\footnote{A preliminary version of this work appears in Proceedings of the 6th Workshop on Coordination of Decentralized Finance (CoDecFin) 2025.}}


 \author{ Mohammed Ababneh, Kartick Kolachala, Roopa Vishwanathan}
 \affil{New Mexico State University, Las Cruces, NM, USA}
 \affil{\href{mailto:mababneh@nmsu.edu@nmsu.edu}{mababneh@nmsu.edu},
 \href{mailto:kart1712@nmsu.edu}{kart1712@nmsu.edu},  \href{mailto:roopav@nmsu.edu}{roopav@nmsu.edu}}

%
\maketitle
\begin{abstract}
Payment channel networks (PCNs) are a promising solution to address blockchain scalability and throughput challenges, However, the security of PCNs and their vulnerability to attacks are not sufficiently studied. In this paper, we introduce $\ulock$, a framework that includes two novel congestion attacks on PCNs. These attacks consider the minimum transferable amount along a path (path capacity) and the number of channels involved (path length), formulated as linear optimization problems. The first attack allocates the attacker's budget to achieve a specific congestion threshold, while the second maximizes congestion under budget constraints.
Simulation results show the effectiveness of the proposed attack formulations in comparison to other attack strategies. Specifically, the results indicate that the first attack provides around a 40\% improvement in congestion performance, while the second attack offers approximately a 50\% improvement in comparison to the state-of-the-art. Moreover, in terms of payment to congestion efficiency, the first attack is about 60\% more efficient, and the second attack is around 90\% more efficient in comparison to state-of-the-art. 
\end{abstract}
\section{Introduction}
\label{sec:intro}

Cryptocurrencies, the most popular blockchain-based application, enable users to transfer money securely and efficiently without relying on centralized authorities such as banks or governments. However, despite their increasing popularity, scalability remains a significant challenge. 
 For instance, Bitcoin blockchain generates a  1MB block every 10 minutes, processing only 7 transactions per second, while users have to wait around one hour (i.e., six blocks) for the transaction to be confirmed~\cite{li2020decentralized}. 
 In contrast, Visa handles 24,000 transactions per second \cite{Visa}.
 
 Payment channel networks (PCN)~\cite{paymentchannels} have emerged as an off-chain solution to the scalability challenge, such as Bitcoin's Lightning Network (LN)~\cite{poon2016bitcoin} and Ethereum's Raiden Network~\cite{Raiden}. The main component of a PCN is a payment channel, where two nodes establish a channel to conduct transactions. Nodes not directly connected with a channel can route transactions through intermediate nodes using multiple hops, extending the network and enabling transactions across the entire PCN. One of the main advantages is that PCNs do not require access to the blockchain every time a transaction occurs (unless there is a dispute). Additionally, PCNs do not require any additional consensus mechanism other than the one employed by the underlying blockchain, and do not require complex cryptographic machinery to process transactions.  This offers the advantage of enabling transactions to be executed over the PCN with relatively low latency and high throughput.

Despite the advantages mentioned above, PCNs are 
vulnerable to attacks. Several types of attacks have been proposed such as griefing~\cite{htlcharm}, LockDown~\cite{perez2020lockdown}, congestion~\cite{lu2020general,mizrahi2021congestion}, wormhole attack \cite{malavolta2018anonymous}, Flood and Loot attack \cite{harris2020flood} and more. 
The common goal of these attacks is to
exploit the inherent limitations and characteristics of the PCN to either gain financial advantage, collect information about
the network, or disrupt its operation. Thus, these attacks can harm the
network’s efficiency, reliability, and user experience. 
Nonetheless, it is important to study and propose new attacks in PCNs as they are useful in drawing attention to network weaknesses and limitations. It will prompt researchers to design corresponding
attack mitigation techniques, thus improving overall network security.

This paper focuses on congestion-based attacks, in which the attacker node issues a payment contract with a specific payment amount to another attacker node over paths involving intermediate nodes.\footnote{As a part of ethical considerations for vulnerability disclosure, we have informed Lightning Labs \cite{lightninglabs} about the attacks described in this paper. Lightning Labs is currently reviewing our attacks and the corresponding results. We also present a few mitigation strategies for the attacks in the paper.}     As the attacker manages both the sender and receiver of the payments, they can intentionally postpone executing the payments, intensifying the congestion in PCNs. These attacks aim to congest payment paths (constituting channels), reducing the network's ability to process legitimate further payments.
While previous studies have explored various aspects of congestion attacks, an important consideration that needs to be considered is congestion attack efficiency. In this work, we aim to address this gap by examining how the attackers can optimize their strategies to congest the network more effectively, given limited resources. In particular,
we investigate how, in PCN, multiple attacker nodes with predefined budgets can allocate path payments to congest the network efficiently.

To address this problem, several factors must be considered, including the attacker's resources, i.e., attack budget, payment path information, e.g., lengths, capacity, and congestion performance. We propose several metrics to quantify the congestion performance more effectively. Using these metrics, we formulate congestion attacks as linear optimization problems. In the first formulation, our goal is to minimize the total amount of payments allocated by the attacker over the different available paths to achieve a threshold congestion performance.  In the second formulation, given an attack budget, the attacker aims to maximize the congestion performance over the different payment paths. In essence, both problems aim to provide a mathematical formulation of efficient resource-limited congestion attacks. To summarize, the contributions of our work are as follows:

\begin{enumerate}
\item We introduce novel metrics that accurately quantify the impact and effectiveness of congestion attacks on PCNs. These metrics provide a more precise evaluation of congestion severity across multiple dimensions, including channel capacity and path length.
\item We present $\ulock$, a comprehensive framework that includes two innovative congestion attack strategies on PCNs. These attacks are formulated as linear optimization problems, providing a mathematical framework for designing efficient, resource-limited congestion attacks. Attackers can either minimize budget expenditure while achieving a desired congestion threshold or maximize the overall congestion impact across the network under predefined budget constraints.

\item We evaluate the performance of $\ulock$ through extensive simulations of a PCN. The simulations validate the effectiveness of the proposed congestion attacks compared to existing strategies, demonstrating superior resource efficiency and congestion performance.

\end{enumerate}
\textbf{Outline}: In Section ~\ref{sec:Back}, we discuss necessary preliminary information. In Section~\ref{sec:related}, we discuss the relevant related work. 
In Section \ref{sec:sysmod}, we present our congestion attacks.
In Section \ref{sec:impl}, we present our experimental evaluation. 
In Section \ref{sec:conc} we conclude the paper.

\section{Background}
\label{sec:Back}
\noindent
\noindent

\subsection{Congestion Attack}
\label{sec:congestion}
The $\lightning$ is susceptible to congestion attacks~\cite{mizrahi2021congestion,lu2020general}. In a congestion attack, the attacker adds Sybil nodes to PCN by establishing a payment channel with existing nodes. Subsequently, The attacker initiates numerous simultaneous payments across multiple paths to either lock capacity along all paths or initiates tiny payments to occupy available HTLC slots along a path and withholds the preimages of the payments between the Sybil nodes. This locks coins across paths, preventing intermediate nodes from earning fees and disrupting their operations.  
Congestion attacks are cost-effective for the attacker. The attacker incurs on-chain fees for opening and closing channels but avoids routing fees as payments remain incomplete. The congestion attacker aims to reduce LN throughput, cause
transaction failures, and eliminate competition by blocking competing nodes and
redirecting traffic and fees to their own nodes.


\begin{figure*}[h!]
    \centering
    \begin{subfigure}[b]{0.49\textwidth}
        \centering
        \includegraphics[width=\textwidth]{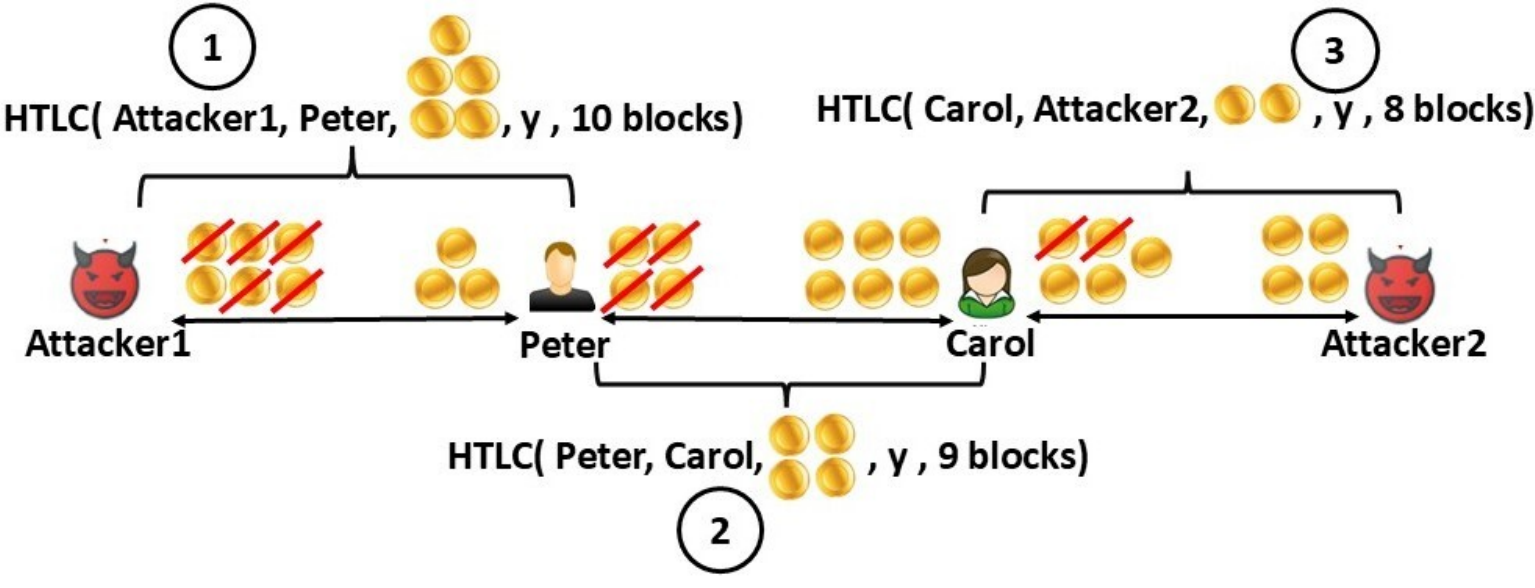}
        \caption{}
        \label{fig:congestionattacke1}
    \end{subfigure}
    \hfill
    \begin{subfigure}[b]{0.49\textwidth}
        \centering
        \includegraphics[width=\textwidth]{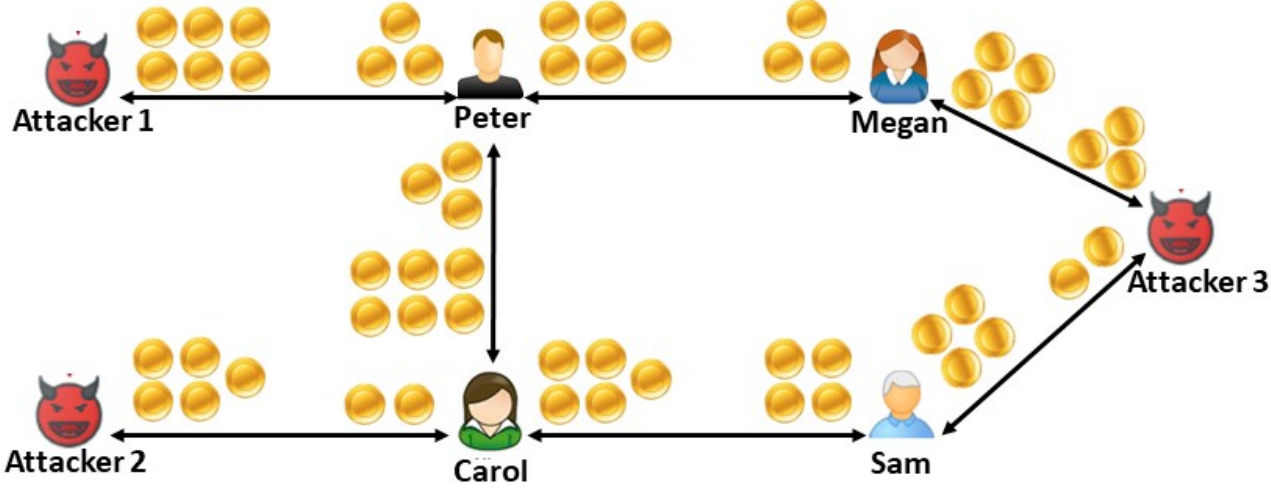}
        \caption{}
        \label{fig:congestionattack}
    \end{subfigure}
    \caption{(a) Channel Capacity Limitation. (b) Congestion Attack}
\end{figure*}

In the congestion attack shown in Figure \ref{fig:congestionattack}, we have three Sybil nodes
($\attackerone$, $\attackertwo$, and $\attackerthree$). For simplicity, we identified a subset of paths between these Sybil nodes are identified as follows:
 $\attackerone$ $\rightarrow$ $\peter$ $\rightarrow$ $\megan$ $\rightarrow$ $\attackerthree$, $\attackerone$ $\rightarrow$ $\peter$ $\rightarrow$ $\carol$ $\rightarrow$ 
$\attackertwo$,
 $\attackertwo$ $\rightarrow$ $\carol$ $\rightarrow$ $\sam$ $\rightarrow$ $\attackerthree$, and  $\attackerthree$ $\rightarrow$ $\sam$ $\rightarrow$ $\carol$ $\rightarrow$ $\attackertwo$. Each attacker generates a payment and sends it along its respective path, with the receiving attacker intentionally withholding the payment execution. In the case where two paths share a common payment channel, such as between $\attackerone$ $\rightarrow$ $\peter$ $\rightarrow$ $\carol$ $\rightarrow$ 
$\attackertwo$, and $\attackerthree$ $\rightarrow$ $\sam$ $\rightarrow$ $\carol$ $\rightarrow$ $\attackertwo$, which both share the payment channel $\carol$ $\rightarrow$ $\attackertwo$, accurate channel usage is critical. To address this, each path is evaluated sequentially, and the capacities of the shared channels are updated after each evaluation. For example, if Attacker1 initiates payments over its paths, the capacity of shared channels, such as Carol → Attacker2, is updated to reflect the locked amounts before subsequent paths are considered. This approach ensures that no two paths simultaneously attempt to use the full capacity of a shared channel, thereby avoiding resource conflicts. The
Hashed Time Locked Contracts (HTLCs) and their Limitations are described in
in Appendeix~\ref{HTLClimtation}. The implications of congestion attacks on PCNs are described in the Appendix.~\ref{implication}.

\section{Related Work}
\label{sec:related}

In this section, we analyze various studies examining attacks on PCNs. Our review focuses on congestion and other related attacks, exploring how attackers can exploit network limitations, such as channel capacity and maximum accepted HTLCs, to disrupt payment flows and compromise network security. Drawing on our insights and observations from prior works, $\ulock$ proposes new congestion attacks and introduces new congestion attack metrics. 
Mizrahi and Zohar~\cite{mizrahi2021congestion} proposed the first congestion attack where 
the attacker uses a greedy algorithm to construct the paths
by consecutively picking high-weighted edges (using channel capacity weight or betweenness centrality). 
Betweenness centrality measures the number of shortest paths passing through the edge, indicating its importance in the PCN. The attacker generates dust payments and sends them over a given path to hold all the available HTLCs, preventing the payment channel along the path from processing more payment transactions. The attacker targets one path at a time, but by repeating the attack across different paths, multiple paths eventually can be congested. In another work,  Lu \emph{et al}.~\cite{lu2020general} proposed a general congestion attack, where the attacker generates Sybil nodes and connects them to strategically chosen set of nodes, typically, nodes with high channel capacities to route more griefing payments. The attacker finds paths between all pairs of Sybil nodes, 
and employs the Ford–Fulkerson algorithm to find the maximum flow over each path. The attacker generates numerous payments and sends them over all paths simultaneously. This causes network congestion and disruption of network performance. In ~\cite{htlcharm}, the authors proposed a griefing attack, where the attacker refuses to respond to an HTLC, which locks all funds along a path until the HTLC time expires. 
Pérez-Sola \emph{et al.}~\cite{perez2020lockdown} presented a lockdown attack that blocks an intermediate node in the multiple-path payment by depleting the capacity in all its channels, which affects its ability to participate in payment routing.  
Malavolta \emph{et al.}~\cite{malavolta2018anonymous} proposed a wormhole attack in which two intermediate nodes in a payment path collude to exclude other intermediate nodes from participating in the payment, thereby stealing the fees intended for those nodes.  
 Harris and Zohar~\cite{harris2020flood} proposed an attack called Flood \& Loot, where the attacker controls source and receiver nodes. The source node initiates payments but denies HTLC fulfillment, forcing channel closures and on-chain claims.
 Jourenko \emph{et al.} proposed Payment Trees \cite{jourenko2021payment}, a payment protocol for PCNs. In this work, the authors construct a virtual channel between the sender and the receiver (on top of already existing payment channels), such that the intermediate node(s) along the payment path lock collateral. Any malicious behavior by the intermediate nodes, such as causing congestion or denial of service is punishable since the malicious nodes lose their collateral. This protocol requires modifications to the current mechanism in which off-chain payments are processed in $\lightning$ and hence is not compatible  with $\lightning$. In $\ulock$, the payment mechanism we use for demonstrating congestion does not require any modifications for payment processing in $\lightning$.



In contrast to other methods, our approach, $\ulock$, frames congestion as linear optimization problems, providing a more precise and resource-efficient attack strategy. 
This sets $\ulock$ apart from existing approaches by optimally allocating resources and targeting congestion at the path level rather than relying on heuristic or greedy algorithms~\cite{mizrahi2021congestion,lu2020general}. $\ulock$ introduces new metrics that consider channel capacity and path length, allowing for a broader attack surface encompassing multiple dimensions of the network’s structure. 



\section{System Construction}
\label{sec:sysmod}
In this section, we describe the assumptions and provide the necessary notations and definitions for the PCN and the attacker. We then present the metrics used to evaluate the effectiveness of congestion attacks. 
Additionally, we examine the implications of congestion attacks on PCNs and propose mitigation techniques to address the congestion attacks.



\subsection{Network Model and Assumptions}
\label{sec:model}
A PCN can be modeled as a directed graph $G=(V,E)$ in which the set of vertices $V$ represent network nodes (i.e., users) and $E$ represents the set of directed edges (i.e., payment channels) between nodes. Assume nodes $v_{x}$, $v_{y}$ $\in$ $V$ have established a payment channel $e_{x,y}$ $\in$ $E$ between them. Let $b_{x,y}$ represent the balance from node $x$ to $y$ (i.e., how many coins can node $x$ forward in the direction of node $y$). Note that the balances $b_{x,y}$ and $b_{y,x}$ are not necessarily equal. The capacity of the channel $e_{x,y}$ is denoted as $c_{x,y}$, where $c_{x,y}$ = $b_{x,y}$ + $b_{y,x}$.


The attacker deploys a set of $N$ Sybil nodes denoted as $\mathcal{A}=\{{A_{1}, A_{2}, \ldots, A_{N}}\}$. The attacker nodes can attach to the PCN nodes using a number of different attachment strategies (e.g., random, highest degree, etc.). The attacks in $\ulock$ can be carried out regardless of the strategy. We now outline our key assumptions that enable the attacks. First, the attacker communicates with the Sybil nodes out-of-band, and any information gathered by the Sybil nodes is accessible to the attacker. Second, the attacker is aware of the entire network topology, including all of the payment channels (i.e., network graph) and their initial capacities and balances. This knowledge can be attained using probing protocols such as~\cite{mizrahi2021congestion}. The probing path does not necessarily involve the network; it is sufficient to prob a number of paths between the attacker pairs. This enables performing a timely attack in which the channel capacities do not change significantly. Assuming that channel capacities undergo a significant change, attack payment will fail. The attacker node (sender) knows immediately that the change happened and thus can either start another prob or reduce the attack payment by updating values of the constraint in the optimization problem.

\subsection{Congestion Attack Metrics}
\label{sec:metrics}
We now propose and describe metrics used to evaluate the effectiveness of $\ulock$'s congestion attacks. In particular, these metrics quantify the impact of the attack on the PCN. Our metrics are: i) Channel Congestion Ratio (CCR), ii) Path Congestion Ratio (PCR), iii) Scale Path Congestion Ratio (SPCR).



To define the metrics, consider two arbitrary Sybil nodes $A_{n}$ and $A_{m}$ and let $P_{j}$ denote the $j^{th}$ path (whose length is measured in terms of number of constituent channels) between them with the attack originating from node $A_{n}$. let $\mathcal{P}=\{{P_{1}, P_{2}, \ldots, P_{J}}\}$ represent all of the paths between all of the pairs of attackers $A_{n}$ and $A_m$.
Furthermore, let $e_{j}^{i}$ denote the $i^{th}$ channel along the path $P_j$. We assume $A_{n}$ runs a probing algorithm. Probing is a technique used by attackers to determine the balance of a payment channel. This is accomplished by sending probes, essentially fake payments whose HTLC digest(s) do not correspond to a valid preimage~\cite{alexprobing,tikhomirov2020probing}. Let us denote the corresponding channel capacity of channel $i$ by $c_{j}^{i}$, and the estimated channel balance along the path $P_{j}$ as $b_{j}^{i}$. Finally, let $\alpha_{j}^{n,m}$ denote the congestion payment made along path $P_{j}$ from attacker $A_n$ towards attaKer $A_m$. Having introduced the above quantities, we now define our performance metrics.

\begin{definition}
\textbf{Channel Congestion Ratio (CCR)} is the ratio between the  payment (originating from attacker $A_{n}$ to attacker $A_m$ ) $\alpha_{j}^{{n,m}}$  along the path $P_{j}$ and the balance of the $i$-th channel along that path (i.e., $b_{j}^{i}$) 

\begin{equation}\label{eq:1}
CCR_{\alpha_{j}^{n, m}}^{n,m}=
    \begin{cases}
        \alpha_{j}^{{n, m}}/b_{j}^{i} & \text{if } \alpha_{j}^{{n, m}} < b_{j}^{i} \\
        1 & \text{if } \alpha_{j}^{{n, m}} = b_{j}^{i} \\
    \end{cases}
\end{equation}

\end{definition}

The CCR quantifies the congestion effect of assigning a payment $\alpha_{j}^{n,m}$ to the channel $e_{j}^{i}$ to process further payments. Naturally, the larger the locked payment $\alpha_{j}^{n,m}$ relative to the balance $b_{j}^{i}$, the more congested a channel is and the larger is the channel's CCR. A channel is fully congested when $\alpha_{j}^{n,m}$ equals its balance and therefore is assigned a value of $1$. 

Note that the CCR quantifies congestion over a single channel along a path. A more comprehensive view of the congestion attack can be provided by considering payment paths instead of individual channels. 
When considering congestion along a path, the congestion bottleneck is determined by the channel with the minimum balance $b_{j}^{i}$.
Regardless of the balances of other channels along a path, the
amount of payment along that path cannot exceed the minimum
channel’s balance.
Thus, the path congestion ratio (PCR) metric is defined as follows.
\begin{definition}
  \textbf{Path Congestion Ratio (PCR)} is defined as the ratio between the payment $\alpha_{j}^{{n, m}}$ and $\min(b_{j}^{i})$ for a path $P_{j}$ and is given as.
  
  \begin{equation}\label{eq:1}
PCR_{\alpha_{j}^{{n, m}}}^{n,m}=
    \begin{cases}
        \alpha_{j}^{{n, m}}/\min({b_{j}^{i}}) &    
        \text{if } \alpha_{j}^{{n, m}}<\min({b_{j}^{i}})  \\
        1 &   \text{if } \alpha_{j}^{{n, m}} = \min({b_{j}^{i}}) \\
    \end{cases}
\end{equation}
  
\end{definition}
Note that the PCR definition above incorporates the bottleneck, which is defined by the minimum balance along the path (i.e., $\min{b_{j}^{i}}$).
Similar to the CCR metric, once the payment  $\alpha_{j}^{n,m}$ equals the $\min({b_{j}^{i}})$, the PCR takes a value of $1$ and the path becomes fully congested. Finally, note that the PCR can be expressed in terms of the CCR as follows:
 \begin{equation}\label{eq:3}
PCR_{\alpha_{j}^{{n, m}}}^{n,m}=
    \begin{cases}
        \max(CCR_{\alpha_{j}^{{n, m}}}^{n,m}) & \text{if } CCR_{\alpha_{j}^{{n, m}}}^{n,m} < 1\\
        1 &  \text{if }  CCR_{\alpha_{j}^{{n, m}}}^{n,m} = 1 \\
    \end{cases}
\end{equation}
  


  
  

One limitation of the PCR definition is that it does not take the path's length into consideration. Paths differ in their lengths  where a path length is defined as the number of channels (i.e., edges) along the path. In addition to the PCR metric, it is informative to consider path lengths as well. Intuitively, a congested path with some PCR value and of length $M$  has a more disastrous  effect on the network than a different path with the same PCR but with a length of $N$ where $M>N$. This is due to the fact that when a payment amount $\alpha_{j}$ is locked along a path $P_{j}$, it is locked on all channels along that path. Thus, in order to incorporate the path length, the scale path congestion ratio (SPCR) is defined next.

\begin{definition}
  \textbf{Scale Path Congestion Ratio (SPCR)}.  
  Let $l_{j}$ denote the length of the $j^{th}$ path $P_{j}$ and $L_{\mbox{\tiny max}}$ refer to the maximum length of a network's path. Then, a path's SPCR is defined as the product of the path's length ratio and its PCR and is given as
 
  \begin{equation} \label{eq:4}
  SPCR_{\alpha_{j}^{{n, m}}}^{^{n,m}}=(\frac{l_{j}}{L_{\mbox{\tiny max}}}) PCR_{\alpha_{j}^{{n, m}}}^{^{n,m}}
\end{equation}

\end{definition}

We note that SPCR incorporates the number of channels congested along a path (i.e., path length) with the path's PCR. Thus, the larger a path's PCR and length, the greater the number of congested channels, resulting in a higher SPCR.The SPCR provides insight into the relative congestion effect of a path compared to others in the network. A higher SPCR corresponds to a longer path, meaning more intermediate nodes along the path are required to lock their coins for the duration of the payment. This introduces the cumulative congestion effect: as coins are restricted across multiple nodes, the network’s ability to handle other transactions decreases. Importantly, it does not matter which specific hop along the path is the bottleneck; the critical point is that all nodes along the long path must reserve the current maximum amount required for the payment. Additionally, SPCR is a directional value that depends on the sender and receiver attack pairs and the origin of the payment. The value of $L_{max}$ determined by the PCN sets the maximum allowable path length. For example, in $\lightning$, this value is set at $20$ hops~\cite{minpath}. Additionally, the SPCR varies between 0 and 1, as discussed earlier, a higher SPCR value indicates a more effective congestion attack.


\subsection{Overview Of Proposed Congestion Attacks}
In this section, we formulate two congestion attacks as linear optimization problems based on the earlier metrics. 
 The first formulation aims to minimize the attacker's total payment  while achieving a desired congestion threshold quantified by the SPCR metric presented earlier. 
 The second formulation aims to maximize congestion effectiveness by maximizing the SPCR while not exceeding the attacker's allowed budget. 
 The SPCR metric plays a critical role in both formulations. It provides a precise measure of congestion attacks by considering both path length and bottleneck capacity. This combination allows for evaluating the detrimental effects that $\ulock$'s congestion attacks have on the 
 PCN, which makes SPCR a valuable metric for optimizing the effectiveness of congestion attacks.
 
It is noteworthy that the proposed congestion attacks are linear in both the objective/cost functions as well the imposed constraints. As such, the problems can be solved either analytically or using linear solvers. However, due to the linear and deterministic nature of the problems they lends themselves easily to a linear solver approach. Next, the congestion attacks are presented in detail.

\subsubsection{$\ulock$'s $\RPO$ Attack}($\minpay$)
\label{sec:paymentmin}
In this attack, an attacker is to minimize budget expenditure while achieving a desired congestion threshold.
This threshold is determined by setting a target SPCR value, which can be viewed as a quality of service (QoS) metric from the attacker's point of view. The larger the SPCR threshold is, the larger the congestion payments are, and the longer the congested paths, thus, the more severe the attack is. $\ulock$'s Payment Minimization Attack can be stated as follows.

\subsubsection{$\ulock$'s $\RPO$ Attack Formulation}
 In this formulation, an attacker $A_{n}$ $\in$ $\mathcal{A}$ is to determine the partial payments on the available paths. In particular, given the set of path capacities and the budget $B_{n}$ of the attacker $A_{n}$ on each path $P_j$, how should the attacker $A_{n}$ send payments over these paths such that the total forwarding payment is minimized. 
    The problem can be mathematically stated as follows;  
\begin{equation}
\min \sum_{\substack{\forall n,m \\ n \neq m}}^ {|\mathcal{A}|} \sum_{j=1}^{|\mathcal{P}|} \alpha_{j}^{n,m}
\label{eq:objective}
\end{equation}
subject to
\begin{equation}
 \sum_{j=1}^{|\mathcal{P}|} \alpha_{j}^{n,m} \leq B_{n} \quad \forall n,m,j
\label{eq:first_constraint}
\end{equation}
\begin{equation}
0 \leq \alpha_{j}^{n,m} \leq \min(b_{j}^{i}) \quad \forall n,m,i,j
\label{eq:second_constraint}
\end{equation}
\begin{equation}
\text{ThresholdValue} \leq \text{SPCR}_{\alpha_{j}^{{n, m}}}^{n,m} \leq 1 \quad \forall n,m,j
\label{eq:third_constraint}
\end{equation}


Where $\alpha_{j}^{n,m}$ denotes the payment allocated by attacker $A_{n}$ over path $P_{j}$ towards attacker $A_m$, and $SPCR_{\alpha_{j}^{{n, m}}}^{n,m}$ denotes the SPCR value of $P_{j}$ resulting from payment allocations by all attacker nodes. Note that the cost function and constraints in the above problem are linear in the payment $\alpha_{j}^{n, m}$. The cost function combines the total payments $\alpha_{j}^{n,m}$ overall congestion paths $P_j$. The constraint in Eqn.~\ref{eq:first_constraint}  is to enforce that an attacker does not exceed its total budget $B_{n}$. The constraint in Eqn.~\ref{eq:second_constraint} ensures that a path payment does not exceed a path's minimum (i.e., bottleneck) capacity since any larger payment can not be supported over the path. The last constraint in Eqn.~\ref{eq:third_constraint} ensures that the attacker achieves a certain congestion threshold over all paths.  Naturally, the higher the threshold value, the more severe the attack is and the more resources (i.e., payments) that need to be used by the attacker.

 

It is important to note that the primary goal of the above problem is to minimize the total payment made by the attacker while achieving a target SPCR value, quantified by the obtained value of SPCR in Eqn.~\ref{eq:4}. Next, we describe the $\ulock$'s $\PCO$ attack that seeks to maximize congestion in the network.

\subsubsection{$\ulock$'s $\PCO$ Attack}($\spcrmax$)
In the previous attack, our goal was to minimize the attack budget to meet, as best as possible, a desired congestion performance in terms of the SPCR. One can envision a second problem of interest as follows; How can we maximize the SPCR congestion performance given a certain attacks budget. This is the attack strategy we develop next


 \subsubsection{$\ulock$'s $\PCO$ Attack Formulation} 
In this formulation, an attacker $A_{n}$ $\in$ $\mathcal{A}$ is to determine the partial payments on the available paths. In particular, given the set of path capacities and the budget of $B_{n}$ of the attacker $A_{n}$ on each path $P_j$, how should the attacker $A_{n}$ send payments over these paths s such that the congestion is maximized. The problem can be mathematically stated as follows; 
 

\begin{equation}
\max \sum_{\substack{\forall n,m \\ n \neq m}}^{|A|} \sum_{j=1}^{|\mathcal{P}|} \text{SPCR}_{\alpha_{j}^{{n, m}}}^{n,m}
\label{eq:max_objective}
\end{equation}
subject to
\begin{equation}
\sum_{j=1}^{|\mathcal{P}|} \alpha_{j}^{{n,m}} \leq B_{n} \quad \forall n,m,j
\label{eq:allocation_constraint}
\end{equation}
\begin{equation}
0 \leq \alpha_{j}^{{n,m}} \leq \min(b_{j}^{i}) \quad \forall n,m,i,j
\label{eq:second1_constraint} 
\end{equation}
\begin{equation}
0 \leq \text{SPCR}_{\alpha_{j}^{{n, m}}}^{n,m} \leq 1 \quad \forall  n,m,j
\label{eq:spcr_bounds}
\end{equation}

 The goal of the objective function in Eqn. \ref{eq:max_objective} is to maximize the total of
  $SPCR_{\alpha_{j}^{{n, m}}}^{n,m}$ along all paths across every pair of attackers. 
 This objective seeks to enhance the overall efficiency or performance across all paths, taking into account both their relative lengths and how well resources are allocated relative to their capacities.
The constraint in Eqn. \ref{eq:allocation_constraint} is the budget constraint which ensures that the total payments $\alpha_{j}^{{n,m}}$  do not exceed the budget $B_{n}$.
The second constraint in Eqn.\ref{eq:second1_constraint} ensures that congested payment $\alpha_{j}^{{n,m}}$ cannot be negative, and it also ensures that a payment of $\alpha_{j}^{{n,m}}$ can be routed via path $P_j$. The third constraint in Eqn.\ref{eq:spcr_bounds} ensures that the $SPCR_{\alpha_{j}^{{n, m}}}^{n,m}$ is between zero and one. It can not be negative or should not be greater than one, indicating over-resource utilization. 
The protocols that describe the setup phase and the launching of $\ulock$'s congestion attacks are described in the Appendix \ref{sec:Alg2} .

\section{Empirical Evaluation}
\label{sec:impl}

This section investigates the performance of the proposed payment minimization attack ($\minpay$) and congestion maximization attack ($\spcrmax$). 
For comparison, we employ two congestion attacks. These congestion attacks are random congestion attack and general congestion attacks~\cite{lu2020general}. In the random congestion attack, both the attacker's budget and path payment are randomly selected. On the other hand,  the general congestion attack is based on a greedy strategy, where the payments are allocated according to path
capacities. Paths with larger capacities are allocated more than the ones with smaller capacities. Moreover, payments are allocated sequentially until the attacker's entire budget is consumed.

For comparison purposes, we use the metrics defined in Section \ref{sec:metrics}.
The first is the PCR metric used to quantify congestion effect over the lowest capacity channels (i.e., bottleneck) in the different paths.   The second is the SPCR  metric which in addition to channel congestion incorporates path length and is indicative of the congestion effect over a path. The higher the PCR and SPCR values, the more effective an attack is in terms of congestion. Another metric is LockedPayment, which represents the amount of locked payment allocated by the attacker. 
As a metric, the locked payment should be taken in conjunction with the achieved congestion performance. Thus, a merely high or low locked payment by itself is not indicative of the overall performance. To address this, we introduce another metric, the cost-congestion efficiency ratio, denoted as   
$\gamma$. 
$\gamma$ is used to quantify how much payment was allocated to provide a given congestion performance. The lower the $\gamma$ value, the more efficient an attack strategy is. Lastly, we use the distribution of PCR values over the attack paths as information to provide an in-depth view and better visualization of congestion distribution among the different attack paths.


\subsection{$\minpay$ Attack Performance Evaluation}
In the first experiment, the performance of the proposed MinPay attack is measured.  The number of attackers is set at $N=6$ attacker pairs (i.e., a total of 12 Sybil nodes), and we experiment with two budgets: 
$B=75\times 10^{6}$ satoshi and $B=100\times 10^{6}$ satoshi. Channel capacities are random with a mean of $4\times 10^{6}$ satoshi. The mean of the path lengths between attacker nodes is $6$ with a maximum length of $8$ channels. Experiment results are averaged over $200$ iterations.  Herein, the performance of the attack is evaluated using different metrics (e.g., PCR and SPCR) as we vary the required SPCR threshold for different $B$ values. 
\begin{figure}[h]
    \centering
    \begin{subfigure}[b]{0.47\linewidth}
        \centering
        \includegraphics[width=1.1\linewidth]{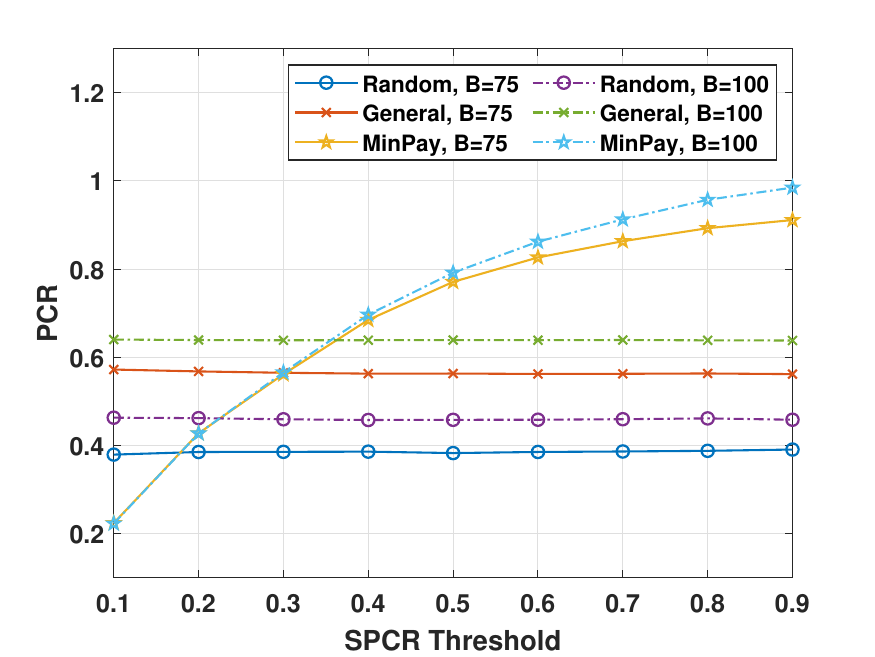}
        \caption{}
        \label{fig:Expr1_PCR_50_75_100}
    \end{subfigure}
    \hfill
     \begin{subfigure}[b]{0.47\linewidth}
        \centering
        \includegraphics[width=1.1\linewidth]{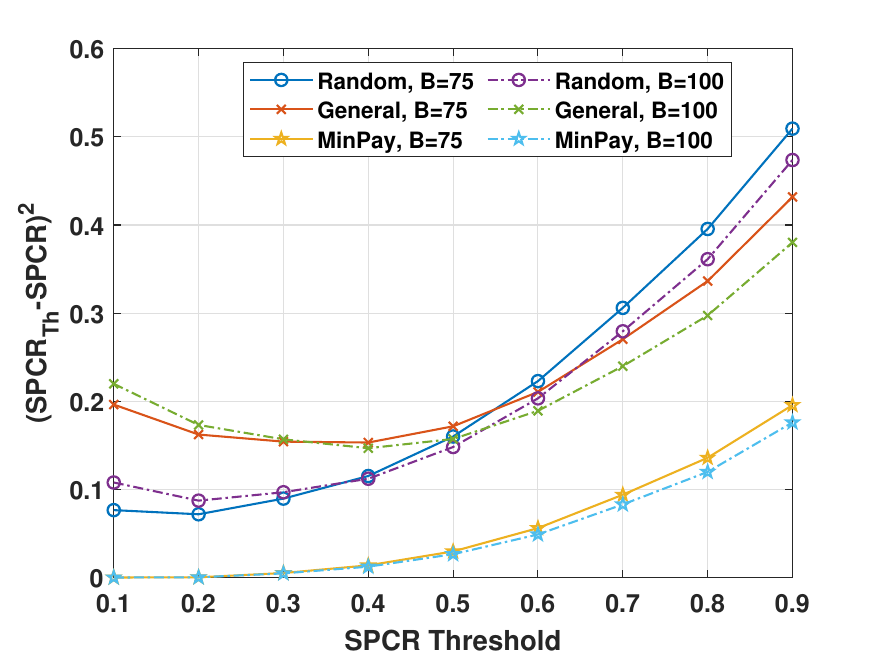}
        \caption{}
        \label{fig:Expr1_SPCR_50_75_100}
    \end{subfigure}
    \caption{(a) PCR vs. SPCR Threshold for different attacker's budget ($B$) values. (b) SPCR deviation for different attacker's budget ($B$) values. }
    \label{fig:costs}
\end{figure}
Results in Fig.~\ref{fig:Expr1_PCR_50_75_100} show the resulting PCR for the different algorithms as a function of SPCR threshold and $B$ values. Note that the PCR of the $\minpay$ attack increases with the SPCR threshold. This is true since as the SPCR requirement increases, it requires more payment to be allocated and thus higher resulting PCR values. We note that for the other algorithms there is no such dependence and thus no change in PCR values. It is also noted that a budget increase from $B=75\times 10^6$ to $B=100 \times 10^6$ results in a noticeable increase in PCR values for $\minpay$ attacks since more payments are available for congestion purposes.

 The SPCR results are shown in Fig .~\ref{fig:Expr1_SPCR_50_75_100} . We note that instead of plotting the achieved SPCR vs.
 SPCR threshold value (denoted as $SPCR_{Th}$, the deviation between achieved and required SPCR is depicted instead. This is because path conditions (e.g., capacity  and length) might make it infeasible to achieve the required SPCR precisely. In this case, and as discussed earlier, the $\minpay$ attack attempts at meeting the SPCR threshold as close as possible with minimum payment. We note that the initial SPCR deviation of the $\minpay$ is close to zero as the threshold is small and thus can be achieved with the given budget. However, as the threshold increases it becomes more difficult to satisfy it and thus the deviation increases. However, we note that the $\minpay$ achieves the smallest deviation in comparison to other attack strategies regardless of the threshold. We also note that the deviation decreases as the budget is increased, which validates our argument.


\begin{figure}[h]
    \centering
    \begin{subfigure}[b]{0.47\linewidth}
        \centering
            \includegraphics[width=1.1\linewidth]{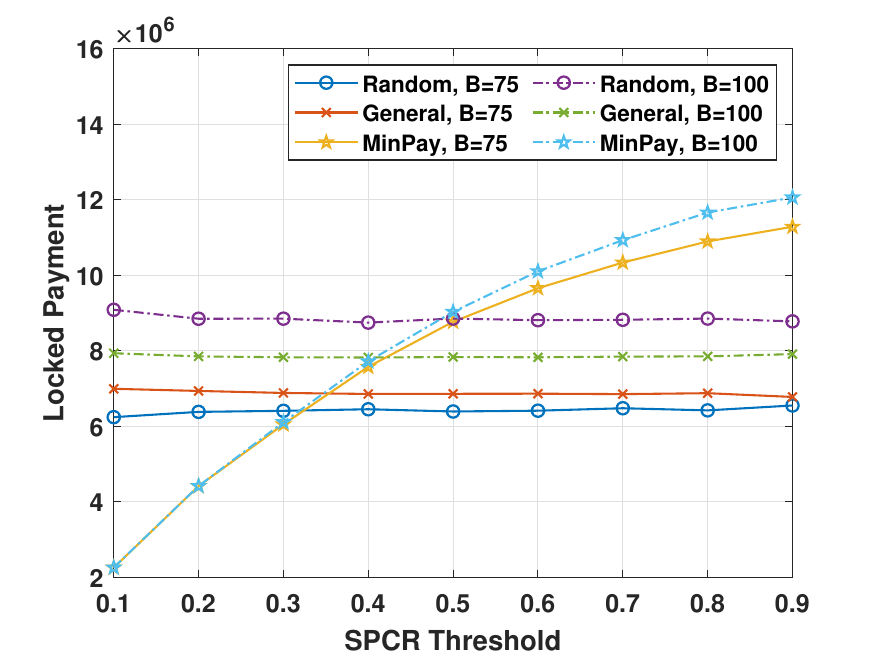}
        \caption{}
        \label{fig:Expr1_Locked_payment}
    \end{subfigure}
    \hfill
     \begin{subfigure}[b]{0.47\linewidth}
        \centering
        \includegraphics[width=1.1\linewidth]{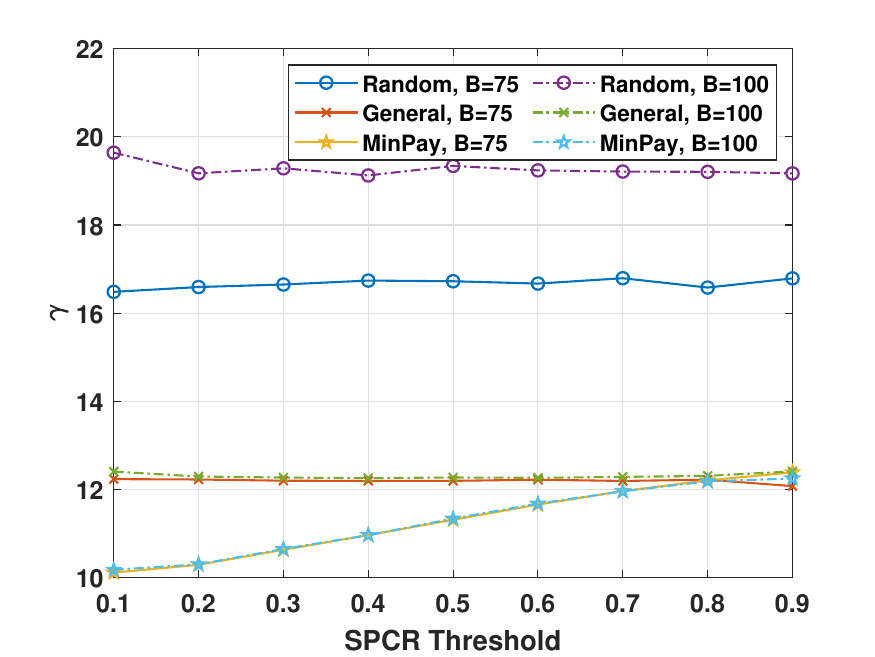}
        \caption{}
        \label{fig:exp1revisedgamma}
    \end{subfigure}
    \hfill
    \caption{(a) LockedPayment vs. SPCR Threshold for different attacker's budget ($B$) values. (b)  \textbf{$\gamma$} vs. SPCR Threshold for different attacker's budget ($B$) values. }
    \label{fig:costs}
\end{figure}

The total payment allocated by the different attacks is summarized in Fig.~\ref{fig:Expr1_Locked_payment}.  However, we note that since the different attacks have varying PCR performances, the locked congestion payment alone is not sufficient for comparison of performances. To this end, we introduce the average cost-to-congestion ratio $\gamma$ defined as: 

\begin{equation*}
	\gamma=\frac{1}{k}\sum_{\substack{\forall n,m \\ n \neq m}}\sum_{j} \frac{\alpha_{j}^{n,m}}{SPCR_{\alpha_{j}^{{n, m}}}^{^{n,m}}}
\end{equation*} 

It is an average operation (i.e., arithmetic or sum of value $\alpha_{j}^{n,m}/SPCR_{\alpha_{j}^{{n, m}}}^{^{n,m}}$ divided over the number of iterations, $k$). A small $\gamma$ value implies a small overall payment and a large PCR value. The more efficient an attack is, the smaller is the $\gamma$ value is.  The  $\gamma$  corresponding to Figs. ~\ref{fig:Expr1_PCR_50_75_100} and ~\ref{fig:Expr1_Locked_payment} is plotted in Fig.~\ref{fig:exp1revisedgamma}. It can be see that the $\minpay$ attack has the lowest  $\gamma$ value and is thus the most efficient. As we can see, The random congestion attack has more variability due to a lack of strategic payment allocation, which makes it sensitive to budget changes from $B=75\times 10^{6}$ satoshi to $B=100\times 10^{6}$ satoshi. 



 \begin{figure*}[h!]
	\centering
	\includegraphics[width=0.51\columnwidth]{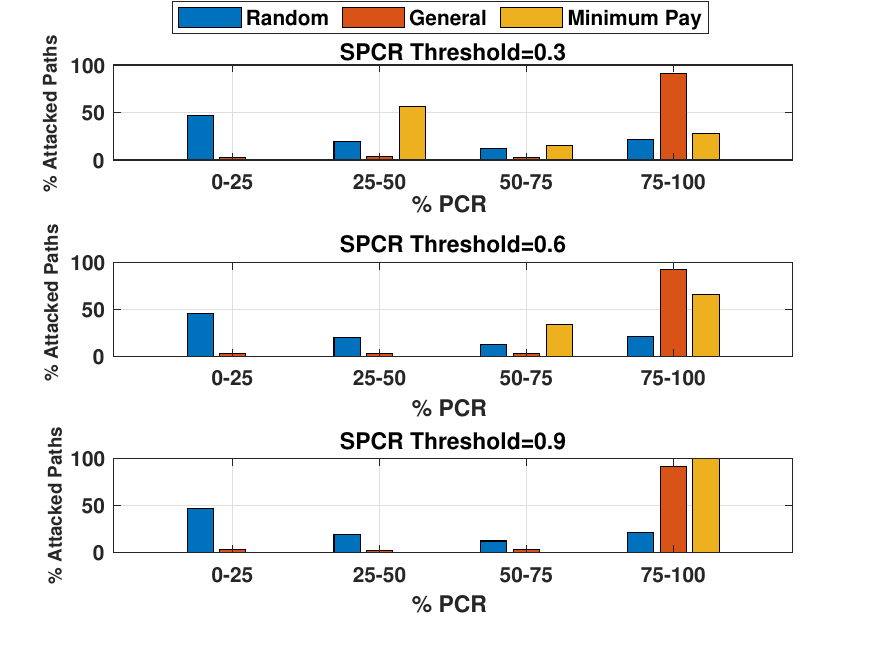}
	\caption[]{$\% PCR$ Distribution}
	\label{fig:Expr1_PCR_Distribution}
\end{figure*}


Results in Fig. ~\ref{fig:Expr1_PCR_Distribution}  show the distribution of PCR values for $B=100\times 10^{6}$ as the SPCR threshold is varied. In particular, the x-axis  represents  PCR values (i.e., congestion ratios) which range from $0-100\%$ divided into $4$ equal intervals. While the y-axis represents  percentage of paths that fall within each PCR interval. We note that achieved PCR distribution for all algorithms, except for the $\minpay$, are similar for all thresholds. However, for the $\minpay$, note that the distribution changes with every SPCR threshold. Note that, with unlimited resources, it is advantageous to have all attack paths be in the 75-100\% interval since this reflects the best congestion performance. However, since in the MinPay it is only required to satisfy the required SPCR threshold value, not all channels fall in the 75-100\% interval. For example, when the threshold is set at 0.3, we note that with the MinPay around 50\% of the paths fall in the 25-50\% interval with the remaining paths having a better congestion (i.e., in the 50-75 and 75-100\%) which is expected. However, as the threshold is increased to 0.9, then 100\% of the paths have a congestion ratio of 75-100\%. This reflects the effectiveness of the MinPay in payment allocation. This is in contrast to other attacks. For example, the random algorithm allocates payments such that most of the channels have a 0-25\% congestion which reflects a poor payment allocation. Moreover, the general attack allocates payments such that most channels are in the 75-100\% range regardless of the threshold. This indicates that inefficient allocation is being made when compared to MinPay.


\subsection{$\spcrmax$  Attack Performance Evaluation}
\begin{figure}[h]
    \centering
    \begin{subfigure}[b]{0.47\linewidth}
        \centering
        \includegraphics[width=1.1\linewidth]{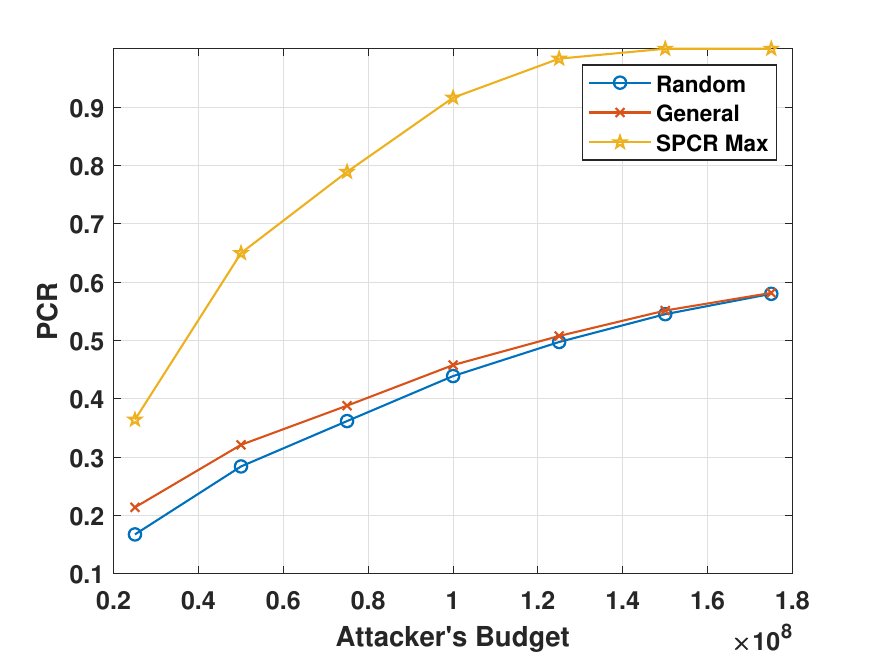}
        \caption{}
        \label{fig:Expr2_PCR}
    \end{subfigure}
    \hfill
     \begin{subfigure}[b]{0.47\linewidth}
        \centering
        \includegraphics[width=1.1\linewidth]{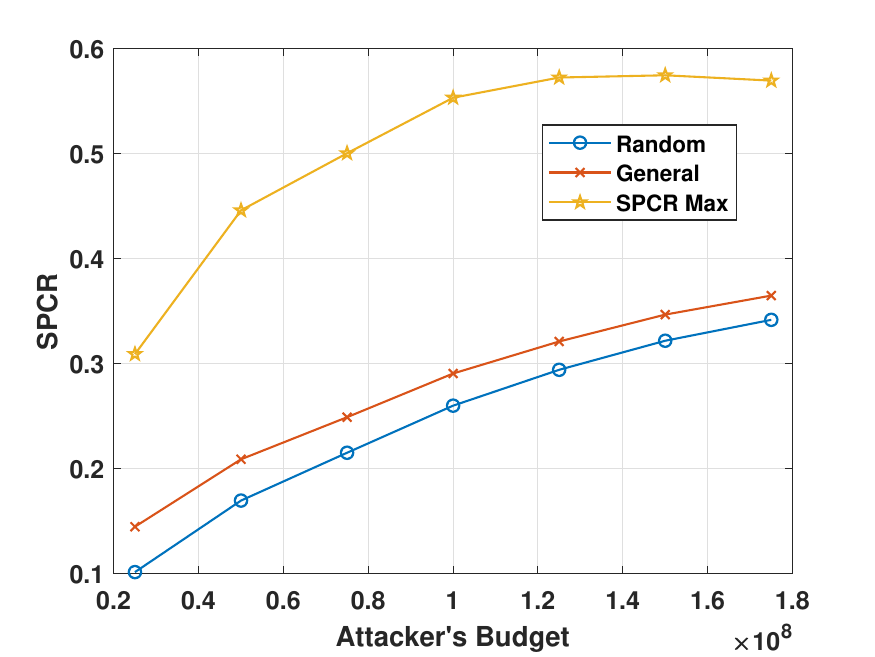}
        \caption{}
        \label{fig:Expr2_SPCR}
    \end{subfigure}
    \caption{(a) PCR vs. Attacker's budget ($B$) values. (b) SPCR vs. Attacker's budget ($B$) values. }
    \label{fig:costs}
\end{figure}

In the second experiment, the performance of the proposed $\spcrmax$ attack is investigated. Here, the goal is to solely maximize the SPCR and thus the resulting PCR as the total allocated budget $B$ is varied. Experiment parameters are the same as the first experiment.
The PCR and SPCR performance of the different attacks are shown in Figs.~\ref{fig:Expr2_PCR} and ~\ref{fig:Expr2_SPCR}.  It is evident that the $\spcrmax$ attack outperforms the other attacks.  Both the PCR and SPCR increase with the attack budget $B$ .  As more budget is available, the PCR reaches the maximum value of $1$ and this results in the saturation of the SPCR as well. In contrast to the PCR, the SPCR value reaches a maximum around $0.57$. This is due to the scaling down of the PCR value by the channel relative length variable of the SPCR (i.e., $l_{P_j}/L_{max}$). Additional simulations show that Lockedpayment vs. Attacker's Budget ($B$) values and  $\gamma$ vs. Attacker's Budget ($B$) values (see Appendix~\ref{ResultsExtended}).
\section{Mitigation Techniques}
\label{sec:Mitigation Techniques}
\textbf{Reputation system.} A reputation-based payment system enhances the security of decentralized networks by imposing initial restrictions on new nodes, limiting payment size and transaction volume to prevent malicious activities like DoS attacks. As nodes build a positive transaction history, these restrictions are gradually relaxed, incentivizing trustworthy behavior. 
\textbf{Updating system parameters.} Adjusting system parameters, such as reducing the maximum payment path length to less than 20 (the max. path length set by $\lightning$~\cite{gossip3}), is one potential strategy to address the issue. However, this approach only postpones the problem instead of solving it, serving as a temporary fix. Additionally, this is a significant change and is unlikely to be implemented in practice. 
\noindent
\textbf{Selectively throttling or rate-limiting of neighbors based on either 1) number of HTLCs, or 2) liquidity.} It can mitigate attacks in the Lightning Network. Parameters such as $htlc\_minimum\_msat$, $max\_accepted\_htlcs$, and $max\_htlc\_value\_in\_flight\_msat$ help nodes limit their exposure by capping the smallest HTLC value, the number of HTLCs, and the total value of outstanding HTLCs, respectively. 
\noindent
 \textbf{Alternate between elephant/mice payments.} This technique refers to how the nodes in PCNs can adjust their willingness to route payments based on current PCN conditions. During periods of high congestion, the nodes may prefer to process mice payments. Otherwise, the nodes will process more elephant payments. This technique ensures consistent throughput in  PCNs. 

\section{Conclusion}
\label{sec:conc}
In this paper, we propose $\ulock$ consisting of two novel congestion attacks, $\PCO$ attack, and $\RPO$ attack, which are presented as linear optimization problems.
The goal of the 
$\RPO$ attack is to minimize budget expenditure while achieving a desired congestion
threshold. The $\PCO$ attack aims to maximize congestion along the paths in PCN ( i.e., making optimal use of an attacker's budget and achieving as much congestion as possible within that budget). The performance of the proposed attacks is compared against random congestion attack and general congestion attack~\cite {lu2020general}.Results show that the proposed attacks outperform existing congestion attacks across multiple metrics. Future work will explore new metrics and attack strategies in PCNs.



\section{Acknowledgments}

This material is based upon work supported by the National Science Foundation under Award No 2148358, 1914635, 2417062,and the Department of Energy under Award No. DESC0023392. Any opinions, findings and conclusions, or recommendations expressed in this material are those of the authors and do not necessarily reflect the views of the National Science Foundation and the Department of Energy.
\bibliographystyle{splncs04}
\bibliography{citation}

\appendix
 \section{Algorithms}
\label{sec:Alg2}
In this section, we describe the algorithms used in the construction of $\ulock$. 
Algorithm~\ref{algo:sybilsetup} describes how the Sybil nodes are generated and how they attach with honest nodes by opening payment channels. First, a set of the honest nodes, $V^{\prime}$, is initialized, where each node generates a key pair using a key generation function for identification. Next, a number of Sybil nodes, $N$, is created, each generating its key pair. A set of targeted nodes, $\tset$, is selected from the honest nodes set, and channels are established between the Sybil nodes and these target nodes. 

\begin{algorithm}[H]
\caption{$\setup$}
\label{algo:sybilsetup}
\tcc{$V^{\prime}$ is the set of honest nodes in the PCN}
\For{$i=j; j\leq |V^{\prime}|; j++$}{
node $i$ performs $\mathsf{KeyGen}(1^\lambda) \rightarrow (\sk_{i}, \vk_{i})$\\
}
\tcc{Sybil nodes being created}
An number of Sybil nodes $N$ is selected\\
$\mathcal{A} = \emptyset$\\
\For{$n=1; n\leq |\mathcal{A}|; n++$}{
    Sybil node $A_n$ is created: $\createsybil(A_n) \rightarrow (\sk_{A_n}, \vk_{A_{n}})$\\
    $\sset = \sset \cup \{\vk_{A_n}\}$\\
}

\tcc{Channel creation between Sybil nodes and honest nodes}
A set $\tset$ of target nodes from $V^{\prime}$ is selected using $\selectnode(V^{\prime}) \rightarrow \tset$, where $|\tset| \geq 1$\\
For each $A_n \in \sset$, a channel is created between $A_{n}$ and  members of $\tset$, $t$ $\in$ $\tset$  using $\pcopen(\vk_{A_{n}}, \vk_{v}, B_{A_n}) \rightarrow success$. \\

\end{algorithm}
 Algorithm~\ref{algo:unified}  implements congestion attacks $\ulock$'s $\RPO$ Attack and $\ulock$'s $\PCO$ Attack, respectively, between pairs of Sybil nodes in the PCN. In both attacks, The attacks begin by iterating over each attacker $A_n \in \sset$. For a given attacker $A_n$, the algorithms find all available paths $\mathcal{P}$ that originate from $A_n$ and connect to all other Sybil nodes $A_{n^\prime} \in \sset$ using  
  using $\sourceroute(A_n, A_{n^\prime})$ function. 
  For each identified path $P_j \in \mathcal{P}$, the algorithms use $\maxflow(P_{j})$ function to determine the minimum channel capacity, $\min(b_{P_j}^{i})$, which represents the bottleneck capacity of the path $P_j$, and computes the hop counts for each path $P_j$ using the $\hopcount(P_{j})$ function, which calculates the number of intermediate nodes along the path $P_j$.These values are stored in the $\Capcitylist$ and $\mathcal{L}$, respectively. In $\ulock$'s $\RPO$ Attack, a threshold value is introduced to minimize the payment costs, ensuring that the allocated payments $\alpha_{P_j}^{{A_n, A_{{n}^\prime}}}$ remain efficient. In contrast, the $\ulock$'s $\PCO$ Attack maximizes congestion across the network without applying a threshold value. After computing the payments $\alpha_{P_j}^{{A_n, A_{{n}^\prime}}}$, the budgets $B_{A_n}$, and the channel capacities $b_{P_j}^{i}$ along all of the paths $P_j$ are updated to reflect the effect of the attack on the network's channels.

  Once the updates are applied, the next attacker $A_n$ begins their iteration. Importantly, this sequential process ensures that each attacker operates on a network state that reflects the cumulative impact of the updates from all previous attackers. As a result, no two attackers see the same network conditions, and the computation of payments $\alpha_{P_j}^{{A_n, A_{{n}^\prime}}}$ adapts dynamically to the updated channel capacities and states. This sequential dependency is critical for effectively simulating and evaluating the evolving impact of the attacks on the network. After all attackers $A_n \in \sset$ have completed their iterations, Step 9 onward executes the congestion attack. This involves creating Hashed Time-Locked Contracts (HTLCs) along each path $P_j$ using the allocated payments $\alpha_{P_j}^{{A_n, A_{{n}^\prime}}}$, propagating requests sequentially between nodes with a timeout $t_{P_j}$ and hash commitment $Y$. The attackers $A_{n^\prime}$ do not execute the payments but use the HTLC setup to congest the network and exploit its resources.
\begin{algorithm}[h!]
\caption{Unified Algorithm for $\ulock$'s $\RPO$ and $\PCO$ Attacks}
\label{algo:unified}

\tcc{This algorithm is run for every pair of Sybil nodes in $\sset$. $L_{max}$ is a system parameter. $\thresval$ is optional and used only in the $\RPO$ Attack.}

Initialize set $\mathcal{P}=\emptyset$  \\
Maintain lists $\Capcitylist=\emptyset$, $\mathcal{L}=\emptyset$, $\mathcal{M}=\emptyset$, $\mathcal{B}=\emptyset$ \\

\For{each attacker $A_n \in \sset$}{
    runs $\sourceroute(A_{n}, A_{n}^\prime)$ to find the set of paths $\pset$ originating from $A_n$ to all other attackers $A_{n}^\prime \in \sset$ \\
    $\forall$ $P_{j} \in \mathcal{P}$, compute $\maxflow(P_{j}) \rightarrow \min(b_{P_j}^{i})$, where $i$ is the index of each channel along path $P_{j}$, and append each $\min(b_{P_j}^{i})$ to $\Capcitylist$ \\
    $\forall$ paths $P_{j} \in \mathcal{P}$, compute $\hopcount(P_{j}) \rightarrow l_{P_j}$, and append $l_{P_j}$ to $\mathcal{L}$ \\
    
    \tcc{$B_{A_n}$ is the budget allocated for attacker $A_n$}
    \uIf{Attack is $\RPO$}{
        Compute $\RPO$ $(\mathbb{C}, \mathcal{L}, B_{A_n}, L_{max}, \thresval) \rightarrow \mathcal{M}$ \tcc*{Uses $\thresval$}
    }
    \ElseIf{Attack is $\PCO$}{
        Compute $\PCO$ $(\mathbb{C}, \mathcal{L}, B_{A_n}, L_{max}) \rightarrow \mathcal{M}$ \tcc*{No $\thresval$}
    }
    
    \tcc{$\mathcal{M}$ includes the attack payments $\alpha_{P_j}^{{A_n, A_{{n}^\prime}}}$ for attacker $A_n$}
    After finding all $\alpha_{P_j}^{{A_n, A_{{n}^\prime}}}$, update $B_{A_n}$ and $b_{P_j}^{i}$ along each path $P_j$
}
\For{each path $P_{j} \in \mathcal{P}$}{
    $A_{n}^\prime$ generates $X \sample \{0,1\}^\lambda$, computes $H(X) \rightarrow Y$, and sends $Y$ to $A_{n}$ \\
    \tcc{ $t_{j}$ is the number of blocks.}
    \For{every pair of consecutive nodes $k, k+1$ along each $P_{j}$}{
        $k$ sends the tuple $\createhtlc(\vk_{k}, \vk_{k+1}, \alpha_{P_j}^{{A_n, A_{{n}^\prime}}}, Y, t_{j})$ to $k+1$ \\
    }
    \If{$curr_{time}$ == $t_{j}$}{
        $k+1$ sends the tuple $\deletehtlc(\vk_{k}, \vk_{k+1}, \mathsf{Cannot~Fulfill})$ to $k$ \\
    }
}
\end{algorithm}

\section{SPCR Max Attack Performance Evaluation}
\label{ResultsExtended}

\begin{figure}[h!]
    \centering
    \begin{subfigure}[b]{0.47\linewidth}
        \centering
        \includegraphics[width=1.1\linewidth]{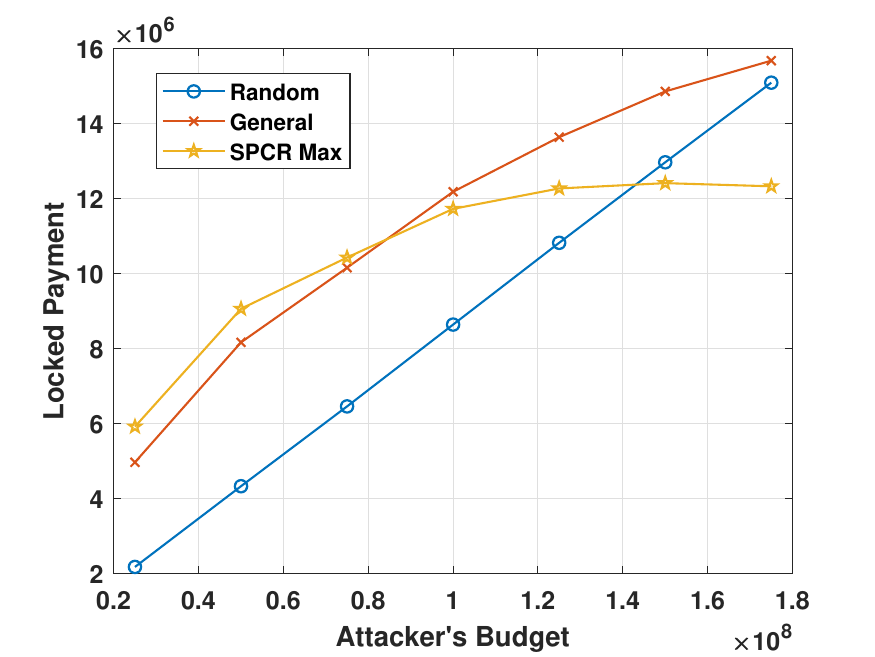}
        \caption{}
        \label{fig:Expr2_Locked_payment}
    \end{subfigure}
    \hfill
     \begin{subfigure}[b]{0.47\linewidth}
        \centering
        \includegraphics[width=1.1\linewidth]{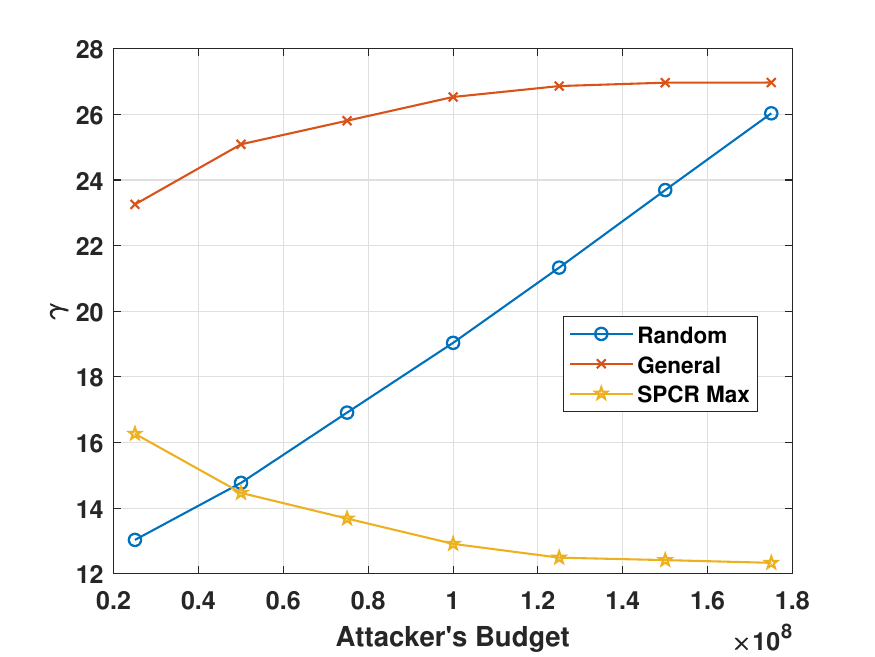}
        \caption{}
        \label{fig:Exp2_revisedgamma}
    \end{subfigure}
    \caption{(a) Lockedpayment vs. Attacker's Budget ($B$) values. (b)  $\gamma$ vs. Attacker's Budget ($B$) values }
    \label{fig:costs}
\end{figure}

The locked payment results are shown in Fig.~\ref{fig:Expr2_Locked_payment} where it is shown that payments allocated by both the random and general attacks increase with $B$. However, the $\spcrmax$ attack does not show the same increase pattern. This is due to the fact that the $\spcrmax$ is able to attain the maximum PCR/SPCR with less allocated funds which results in the plateauing of its curve. The relative efficiency of the different attacks can be quantified using the $\gamma$ parameter which is depicted in Fig.~\ref{fig:Exp2_revisedgamma}  The results indicate that the $\spcrmax$ becomes more efficient as more attack budget is available. In fact, at large $B$ values, it can be more than two times more efficient than other algorithms. 

\section{Implications of  Congestion Attacks on PCNs}
\label{implication}
\begin{enumerate}
    \item Targeting High Liquidity Nodes: The adversary in $\ulock$ can target the nodes with high liquidity and/or those with a large number of payment channels (high degree nodes). These high liquidity/degree nodes are often referred to as routing helpers/trampolines/trampoline nodes \cite{trampolineslides} and play a pivotal role in facilitating transactions across the $\lightning$. By establishing payment channels with these nodes, the adversary can cause significant disruptions by locking their liquidity, preventing them from routing transactions, thereby reducing the overall throughput of the $\lightning$.

\item Disconnecting $\lightning$: Payment paths in LN can use payment channels that act as critical bridges, linking independent clusters of nodes and enabling the flow of payments between them. An attacker using $\ulock$ could could intentionally congest these payment channels, disrupting the payments that depend on them.


\item Node's reputation: Nodes in LN facilitate payment forwarding and generate routing fees. However, they can be vulnerable to congestion attacks, which may damage their reputation if they are unable to process requests, resulting in "temporary channel failure" errors. These errors signify insufficient funds, leading senders to lose trust in the node and opt against future transactions.
\end{enumerate}

\section{Hashed Time Locked Contracts (HTLCs) and their Limitations}
\label{HTLClimtation}
The  $\lightning$ uses Hashed Time Locked Contracts (HTLCs) to enable secure, trustless, multi-hop payments. HTLCs ensure staggered payments, where each node forwards payments while being protected from loss through the blockchain's dispute resolution mechanism. However, HTLCs have limitations that can be exploited for congestion attacks:

\noindent
1) \textbf{Maximum Accepted HTLCs Limitation}: 
Nodes have a limit on the number of HTLCs they can handle (e.g., 483 in $\lightning$\cite{mizrahi2021congestion}). Attackers can fill these slots with numerous small payments (e.g., dust payments of $ 5.46 \times 10^{-6} $ BTC~\cite{mizrahi2021congestion}), blocking legitimate transactions.
2) \textbf{HTLC Expiration Time Limitation}:
HTLCs use a Check Lock Time Verify (CLTV), with 14, 40, or 144 blocks expiring in $\lightning$ implementations. Attackers can exploit this linear accumulation (up to 2016 blocks~\cite{mizrahi2021congestion}) to lock coins for extended periods, delaying transactions.
3) \textbf{Channel Capacity Limitation}:
Channel capacities are fixed at opening, limiting transaction amounts. Attackers can lock coins along paths by withholding preimages, as in Figure~\ref{fig:congestionattacke1},
$\attackerone$ locks five coins with $\peter$, who forwards four coins to $\carol$ after deducting routing fees. $\attackertwo$ then withholds the preimage, locking a total of twelve coins, congesting the path, and reducing path capacity.


%




\end{document}